\documentclass
[preprint,prb,a4paper,amsfonts,amssymb,floats,titlepage,fleqn,footinbib,12pt,onecolumn,nobalancelastpage]{revtex4}%
\usepackage{amsfonts}
\usepackage{amsmath}
\usepackage{amssymb}
\usepackage{graphicx}
\usepackage{longtable}%
\providecommand{\U}[1]{\protect\rule{.1in}{.1in}}

\begin{document}
\author{ M. ElMassalami, R. Moreno}
\affiliation{Instituto de Fisica, Universidade Federal do Rio de Janeiro, Caixa Postal
68528, 21945-970 Rio de Janeiro, Brazil}
\author{H. Takeya}
\affiliation{National Institute for Materials Science,1-2-1 Sengen, Tsukuba, Ibaraki
305-0047, Japan}
\author{B. Ouladdiaf}
\affiliation{Institut Laue-Langevin, B.P.156 ,38042 Grenoble Cedex 9, France}
\author{J. W. Lynn}
\affiliation{NIST Center for Neutron Research, National Institute of Standards and
Technology, Gaithersburg, MD 20899-6102}
\author{R. S. Freitas}
\affiliation{Instituto de F\'{\i}sica., Universidade de S\~{a}o Paulo, Rua do Mat\~{a}o 187
Travessa R, Cidade Universit\'{a}ria, 05315-970 Sao Paulo, SP , Brasil }
\title{Magnetic structures of quaternary intermetallic borocarbides $R$Co$_{2}$%
B$_{2}$C ($R$=Dy, Ho, Er).}
\date{\today{}}

\begin{abstract}
The magnetic structures of the title compounds have been studied by neutron
diffraction. In contrast to the isomorphous $R$Ni$_{2}$B$_{2}$C compounds
wherein a variety of exotic incommensurate modulated structures has been
observed, the magnetic structure of ErCo$_{2}$B$_{2}$C is found to be
collinear antiferromagnet with $k$=($\frac{1}{2},0,\frac{1}{2}$) while that of
HoCo$_{2}$B$_{2}$C and DyCo$_{2}$B$_{2}$C are observed to be simple
ferromagnets. For all studied compounds, the moments are found to be confined
within the basal plane and their magnitudes are in good agreement with the
values obtained from the low-temperature isothermal magnetization
measurements. The absence of modulated magnetic structures in the $R$Co$_{2}%
$B$_{2}$C series (for ErCo$_{2}$B$_{2}$C, verified down to 50 mK) is
attributed to the quenching of the Fermi surface nesting features.

\end{abstract}

\maketitle

\section{Introduction}

One of the most striking features of the magnetism of the $R$\textrm{Ni}$_{2}%
$\textrm{B}$_{2}$\textrm{C }series is the manifestation of a variety of
incommensurate modulated antiferromagnetic-like modes, some of which coexist
with a superconducting ground
state.\cite{Muller01-interplay-review,Lynn97-RNi2B2C-ND-mag-crys-structure,Canfield-RNi2B2C-Hc2-review}
There are three different types of modulations that have been observed:
$\overrightarrow{k}_{1}\mathrm{\simeq(}$0.55, 0, 0)\ as in \textit{R }= Er, Ho
[4.5 $<T<$6.5 K], Tb, and Gd; $\overrightarrow{k}_{2}\mathrm{\simeq}$(0.093,
0.093, 0) as in \textit{R }= Tm; and $\overrightarrow{k}_{3}\mathrm{\simeq}%
$(0, 0, 0.91) as in \textit{R }= Ho [5 $<T<$8 K]. In addition, there are two
commensurate antiferromagnetic (AFM)\ structures having $\overrightarrow
{k}_{4}\mathrm{=}$(0,0,1) as in \textit{R }= Ho, Dy, Pr and $\overrightarrow
{k}_{5}\mathrm{=}$(0.5,0,0.5) as in \textit{R }= Nd. For the case of \textit{R
}= Ho, the $\overrightarrow{k}_{1},\overrightarrow{k}_{3},$ and
$\overrightarrow{k}_{4}$ modes coexist within a narrow range of
temperature.\cite{Lynn97-RNi2B2C-ND-mag-crys-structure} The manifestation of
such a variety of wave vectors\ is not uncommon in intermetallic magnets
wherein the magnetic couplings are mediated by the indirect
Rudermen-Kittel-Kasuya-Yosida (RKKY) interactions:\cite{Coqblin-book} indeed
for the particular case of $R$\textrm{Ni}$_{2}$\textrm{B}$_{2}$\textrm{C}, the
presence of these RKKY interactions is evidenced as a de Gennes scaling of
both the superconducting\ and magnetic transition
temperatures.\cite{Muller01-interplay-review,Canfield-RNi2B2C-Hc2-review}

As is generally the case, the type of the magnetic structure for the
rare-earth moments in $R$\textrm{Ni}$_{2}$\textrm{B}$_{2}$\textrm{C} is
determined by the competition between these RKKY interactions, the crystalline
electric field (CEF) forces, and the classic dipolar
interactions.\cite{Kalatsky98-clock-model,Amici-Thalmeier98-HoNi2B2C,Jensen02-mode-FM-ErNi2B2C}
The RKKY interactions (and thus the magnetic structure) depend partially on
the spatial separation of the $R^{3+}$ ions as well as on the electronic band
structure of these borocarbides. The importance of the latter is highlighted
by the finding that the $\overrightarrow{k}_{1}$ mode is related to
the\ electronic nesting features that give rise to a peak in the generalized
susceptibility.\cite{Rhee95-generalized-susc}

It is worth mentioning that the manifestation of a variety of wave vectors in
the members of the $RT_{2}$\textrm{B}$_{2}$\textrm{C} family is not forbidden
by group theory
arguments.\cite{Wills03-IR-symmetry,Ballou-ouladdiaf06-RepAnalysis,Izyumov-book}
In fact, for these tetragonal borocarbides ($I$4/$mmm$) where the magnetic
moments reside at the Wyckoff site 2$a$, Wills \textit{et al.}%
\cite{Wills03-IR-symmetry} showed that there are fifteen possible wave
vectors, each (or a superposition) of them can be used to describe a distinct
magnetic structure (see Table 3 of Ref.\cite{Wills03-IR-symmetry}). \ As
mentioned above, the stabilization of a specific mode depends critically on
\ the energy balance among the CEF, dipolar, and exchange interactions. It is
interesting to note that this representational analysis predicts the
possibility of a ferromagnetic (FM) mode in this $RT_{2}$\textrm{B}$_{2}%
$\textrm{C} family: \ such a mode has not been observed in $R$\textrm{Ni}%
$_{2}$\textrm{B}$_{2}$\textrm{C} but, as shown in previous reports
\cite{08-Mag-Structure-TmCo2B2C,08-Mag-Structure-TbCo2B2C} as well as in this
work, a FM mode is manifested in a number of $R$\textrm{Co}$_{2}$%
\textrm{B}$_{2}$\textrm{C} compounds.

The electronic band structures of the $RT_{2}$\textrm{B}$_{2}$\textrm{C}
($T$=Ni, Co) family of compounds are very
similar.\cite{Coehoorn94-RNi2B2C-electronic-structure} In addition, both
series have almost the same Sommerfeld specific-heat coefficients.
Furthermore, each isomorphous pair has almost equal lattice parameters and
very much similar $R^{3+}$ single-ion crystalline-electric field
properties.\cite{00-RCo2B2C} Then it is expected that the multi wave-vectors
character to be manifested also in the isomorphous $R$\textrm{Co}$_{2}%
$\textrm{B}$_{2}$\textrm{C} series. Contrary to\ this expectation, the
reported magnetic structures of $R$\textrm{Co}$_{2}$\textrm{B}$_{2}$\textrm{C}
($R$=Tb [Ref.\cite{08-Mag-Structure-TbCo2B2C}],Tm
[Ref.\cite{08-Mag-Structure-TmCo2B2C}]) are found to be ferromagnetic: in
distinct contrast to the modulated spin configuration of the Ni-based
isomophs. As this work shows, the magnetic structures of $R$\textrm{Co}$_{2}%
$\textrm{B}$_{2}$\textrm{C} ($R$\textit{ }= Dy, Ho, Er) are also ferromagnetic
and different from their Ni-based isomorphs.

The difference in the magnetic structure of these isomorphous $RT_{2}%
$\textrm{B}$_{2}$\textrm{C} ($T$=Ni, Co) series is attributed to the
following:\cite{08-Mag-Structure-TmCo2B2C,08-Mag-Structure-TbCo2B2C}
electronic band structure
calculations\cite{Pickett94-electronic-structure,Matthias94-electronic-structure,Coehoorn94-RNi2B2C-electronic-structure,Lee94-electronic-structure}
demonstrated that the density of states at the Fermi level, $N$($E_{F}$), of
both series receives appreciable contribution from the 3$d$ orbitals of the
transition-metal atoms.\ As that the electronic properties of the $RT_{2}%
$\textrm{B}$_{2}$\textrm{C} family can be reasonably well described in terms
of the rigid band model, the substitution of Co atoms (which has a lower
number of the 3$d$ electrons) induces a downward shift in $E_{F}$ but with a
new $N$($E_{F}$) which is almost equal to that of $R$\textrm{Ni}$_{2}%
$\textrm{B}$_{2}$\textrm{C} [compare Figs. 1 and 4 of Ref.
\cite{Coehoorn94-RNi2B2C-electronic-structure}]. Such a shift entails a
different generalized susceptibilities and thus a different character of the
magnetic ground state. In particular, the absence of modulated states in
$R$\textrm{Co}$_{2}$\textrm{B}$_{2}$\textrm{C} is taken to be a strong
indication that the characteristic nesting features at, say, $\overrightarrow
{k}_{1}$ must have been quenched.

\section{Experiment}

99.5\% $^{11}$B enriched polycrystals of $R$\textrm{Co}$_{2}$\textrm{B}$_{2}%
$\textrm{C} were prepared by conventional arc-melt method. Powder
neutron-diffraction measurements on as-prepared samples were carried out at
the Institut Laue-Langevin in Grenoble, France ($\lambda=$2.359 $%
\operatorname{\mathring{A}}%
$, 0.48 $\leq T\leq$ 4.5 K) and the National Institute of Standards and
Technology, USA ( BT-9 triple-axis instrument with a pyrolytic graphite
monochromator and filter, $\lambda=$2.359 $%
\operatorname{\mathring{A}}%
$, 0.05 $\leq T\leq$ 8.0 K, and BT-1 high resolution powder diffractometer
with $\lambda=$2.0787 $%
\operatorname{\mathring{A}}%
$,).\ Due to the experimental difficulties, the neutron diffractograms within
the mK range were measured only for \textrm{ErCo}$_{2}$\textrm{B}$_{2}%
$\textrm{C} (\S \ III.A). Similarly, detailed temperature-dependent
diffractograms were collected only for \textrm{HoCo}$_{2}$\textrm{B}$_{2}%
$\textrm{C} (\S \ III.B).

Powder Rietveld refinements of both crystallographic and magnetic
structures\ were carried out using the \textsc{Fullprof} package of
Rodriguez-Carvajal (L. L. B.).

Magnetization $M(T,H)$ measurements (\S \ III.B-C) were conducted in a
vibrating-sample magnetometer which was operated within the range [0.5 $<T<$
30 K, $H$%
$<$
17 T].
\begin{figure}
[th]
\begin{center}
\includegraphics[
height=4.6873in,
width=3.1912in
]%
{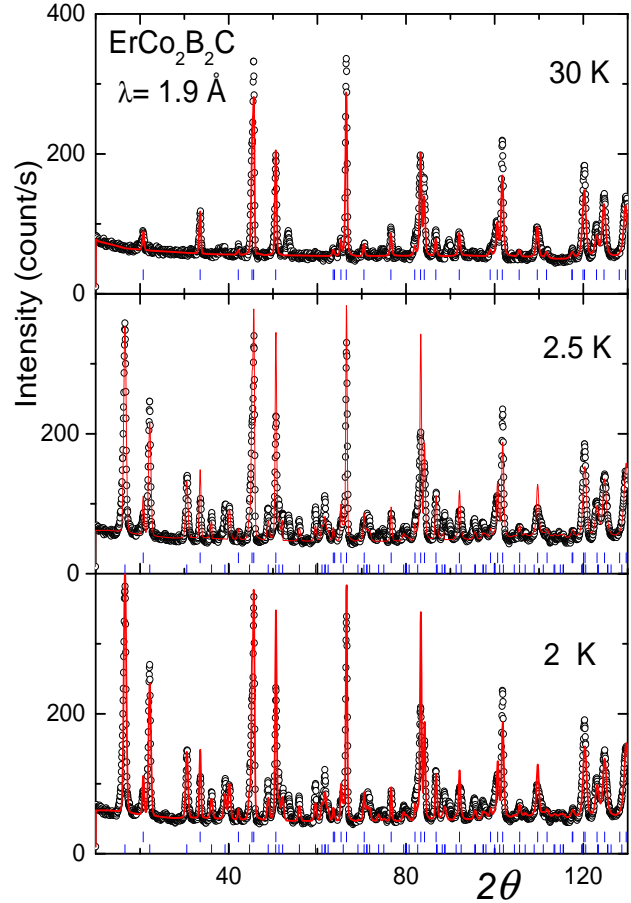}%
\caption{(Color online) Representative neutron diffractograms of ErCo$_{2}%
$B$_{2}$C measured at 2 K (low panel), 2.5 K (middle panel), and 30 K (top
panel). Symbols, solid line, and short vertical bars represent, respectively,
measured intensities, calculated intensities based on Rietveld-refinement, and
positions of the Bragg reflections (see text).}%
\label{erco2b2c-nd-riet}%
\end{center}
\end{figure}

\section{Results}

\subsection{\textrm{E}$\mathrm{r}$\textrm{C}$\mathrm{o}_{2}$\textrm{B}$_{2}%
$\textrm{C}}

\begin{figure}
[ptb]
\begin{center}
\includegraphics[
height=4.6873in,
width=3.2379in
]%
{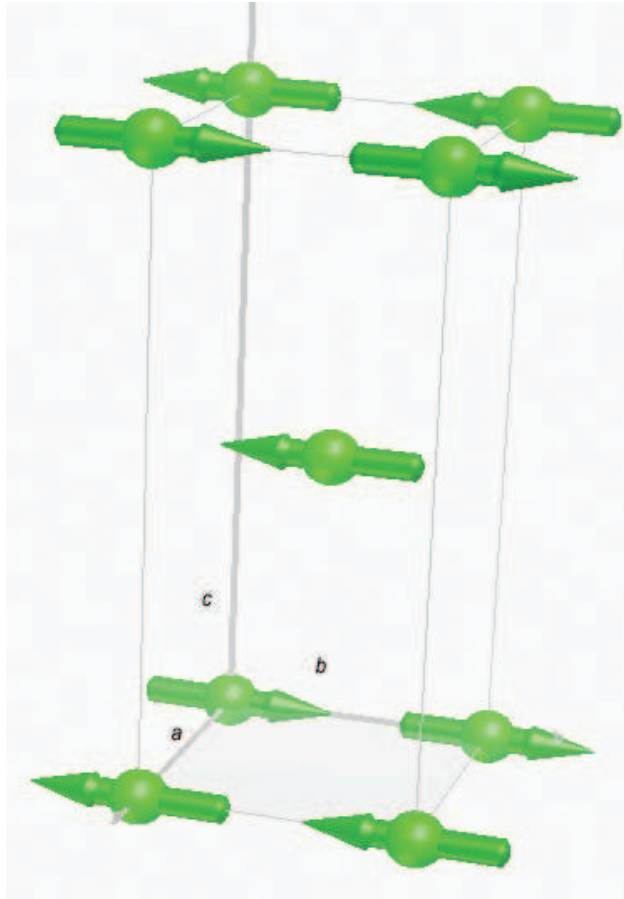}%
\caption{The commensurate, collinear, antiferromagnetic structure of
\textrm{ErCo}$_{2}$\textrm{B}$_{2}$\textrm{C}. The moments are coupled
antiferromagnetically along both the $a$ and $c$ axis and ferromagnetically
along the $b$ axis.}%
\label{erco2b2c-mag-structure}%
\end{center}
\end{figure}

\begin{figure}
[th]
\begin{center}
\includegraphics[
height=4.721in,
width=3.7844in
]%
{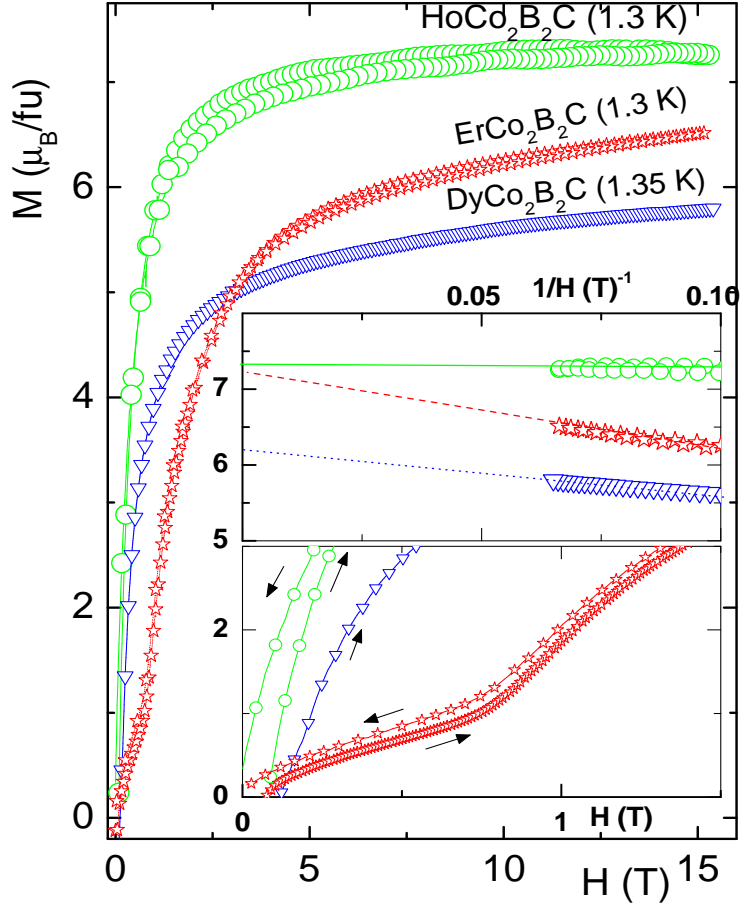}%
\caption{(Color online) Magnetization isotherms of $R$Co$_{2}$B$_{2}$C
($R$=Dy, Ho, Er) measured at the indicated temperatures. The lower inset shows
an expansion of the low-field region of the forced magnetization:
while$\ M(T,H)$ of $R$=Dy, Ho is typical of a forced FM state, that of $R$=Er
is characteristic of a field-induced spin-flop anomaly of an AFM\ structure.
The upper inset demonstrates the saturated magnetization when $H\rightarrow
\infty$ ($\frac{1}{H}\rightarrow0$): for $R$=Er and Dy, no saturation is
attained even for a field of 15 T. The extrapolated values when $\frac{1}%
{H}\rightarrow0$ are shown in Table \ref{Tab.I}. Similar high-field
magnetization features were observed in \textrm{TbCo}$_{2}$\textrm{B}$_{2}%
$\textrm{C }(see Fig. 2 of Ref.\cite{08-Mag-Structure-TbCo2B2C}) and
\textrm{TmCo}$_{2}$\textrm{B}$_{2}$\textrm{C }(see Fig. 6 of
Ref.\cite{08-Mag-Structure-TmCo2B2C}).}%
\label{Fig-rco2b2c-magnetization}%
\end{center}
\end{figure}

Based on earlier zero-field $ac$ susceptibility and specific heat
results,\cite{00-RCo2B2C} the magnetic order of \textrm{ErCo}$_{2}$%
\textrm{B}$_{2}$\textrm{C} sets-in at $T_{N}=4.0$($1$) K. Indeed, below
$T_{N}$ the diffractograms of \textrm{ErCo}$_{2}$\textrm{B}$_{2}$\textrm{C}
(Fig. \ref{erco2b2c-nd-riet}) show a number of additional intense Bragg peaks
which can be indexed on the basis of the\ magnetic structure given in Fig.
\ref{erco2b2c-mag-structure}: a commensurate, collinear, antiferromagnetic
structure with the moments being\ coupled antiferromagnetically along both $a$
and $c$ axes and ferromagnetically along the $b$ axis; i.e. $k$=($\frac{1}{2}$
$0$ $\frac{1}{2}$). This AFM character of the ordered state is manifested also
in the magnetization isotherm of Fig. \ref{Fig-rco2b2c-magnetization}.%


Based on the refinement\ of the powder diffractograms (Fig.
\ref{erco2b2c-nd-riet}), the Er$^{3+}$ moments are found to be confined within
the basal plane and reach 6.8(2) $\mu_{B}$ at 2.0 K. This value is in
excellent agreement with the saturated moment 7.0(2) $\mu_{B}$ obtained from
the isothermal magnetization measured at 1.40(5) K (Fig.
\ref{Fig-rco2b2c-magnetization}). Moreover, it compares favorably with the
value reported for \textrm{ErNi}$_{2}$\textrm{B}$_{2}$\textrm{C} ($\mu
=$7.2$\pm0.$1 $\mu_{B}$),\cite{Lynn97-RNi2B2C-ND-mag-crys-structure}
suggesting a similarity in their single-ion properties. However, there are
important differences between these two isomorphs: the Co-based isomorph does
not superconduct and does not have an incommensurate magnetic structure,
instead ordering into a collinear AFM structure at a critical temperature that
is 40\% lower than that of the Ni-based isomorph.%
\begin{figure}
[th]
\begin{center}
\includegraphics[
height=4.6873in,
width=4.9718in
]%
{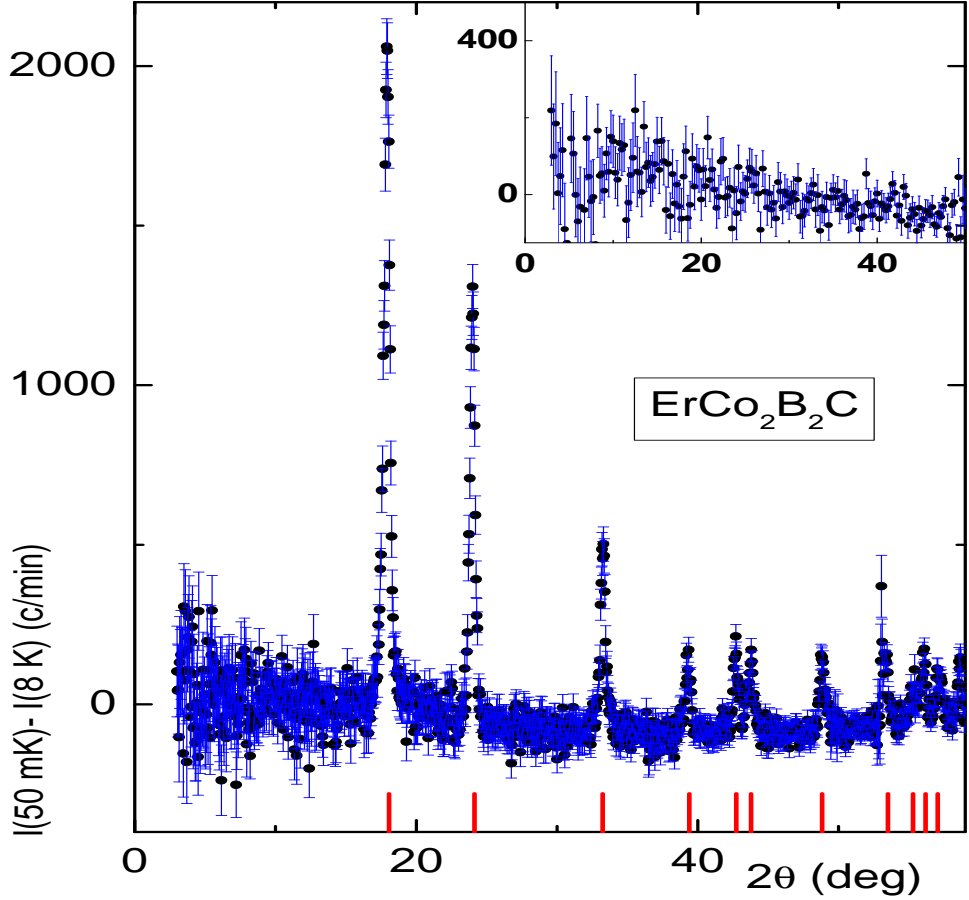}%
\caption{(Color online) Low-temperature magnetic diffractogram of
\textrm{ErCo}$_{2}$\textrm{B}$_{2}$\textrm{C} obtained after subtracting the
nuclear contributions (diffractogram at 8 K) from the one at 50 mK; the solid
vertical bars index the magnetic pattern. The inset shows the difference
pattern obtained after subtracting the diffractogram at 700 mK from that of 50
mK: evidently there is no change in the magnetic structure (see text).
\ Uncertainties where indicated in the manuscript are statistical in nature
and represent one standard deviation.}%
\label{erco2b2c-nd-riet-lowt}%
\end{center}
\end{figure}

It was reported earlier that both the specific heat and $ac$ susceptibility
curves of\ an \textrm{ErCo}$_{2}$\textrm{B}$_{2}$\textrm{C} sample exhibit a
magnetic anomaly at $T_{M}$ = 0.37(2) K.\cite{00-RCo2B2C} For the purpose of
investigating the origin of the reported anomaly, we collected neutron
diffractograms at various temperatures: 50 mK (well below $T_{M}$), 700 mK
(just above $T_{M}$ but well below $T_{N}$), and 8 K (above $T_{N}$).\ As can
been seen in Fig. \ref{erco2b2c-nd-riet-lowt}, there are no additional
magnetic peaks (nor any other features) that can be associated with this
anomaly. In fact the inset of Fig. \ref{erco2b2c-nd-riet-lowt} indicates that
the magnetic patterns at 700 mK and 50 mK are identical. As there are no
change in the diffractograms when the temperature is cooled through the
reported\ $T_{M}$, it is concluded that the reported anomalous transition at
$T_{M}$ may correspond to an ordered moment that is too small to be detected
in powder diffraction, or it does not correspond to a new long-range ordered
state. \ We also cannot rule out that it is a sample dependent effect, which
would imply that it is not intrinsic to the Er-sublattice.

\subsection{\textrm{H}$\mathrm{o}$\textrm{C}$\mathrm{o}_{2}$\textrm{B}$_{2}%
$\textrm{C}}

\begin{figure}
[th]
\begin{center}
\includegraphics[
height=4.6882in,
width=4.7314in
]%
{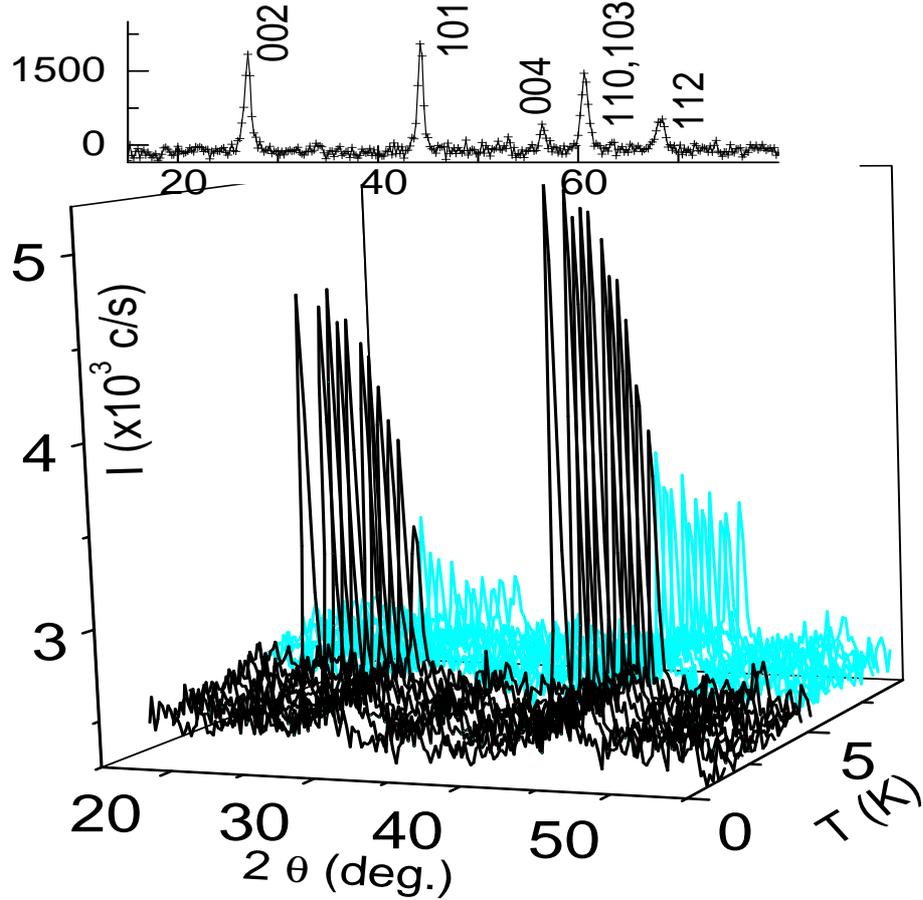}%
\caption{(Color online) Lower panel: three dimensional plot of representative
powder diffractograms of \textrm{HoCo}$_{\mathrm{2}}$\textrm{B}$_{\mathrm{2}}%
$\textrm{C}. The intensities are measured over a wide range of scattering
angle (for clarity, here shown only up to 55${{}^\circ}$) and at various
temperatures. The enhancement of the intensities at the positions of the
nuclear Bragg peaks is clearly manifested below $T_{C}$. The fact that there
are no additional magnetic peaks and that the Ho site is at 4$a$ site
indicates that the magnetic arrangement must be due to an onset of
ferromagnetism (see text). This is confirmed in the upper panel which shows
the difference plot of $I$(1.5K)-$I$(10 K) together with the Bragg peaks
identifications.}%
\label{hoco2b2c-raw}%
\end{center}
\end{figure}

\begin{figure}
[th]
\begin{center}
\includegraphics[
height=4.6882in,
width=3.6703in
]%
{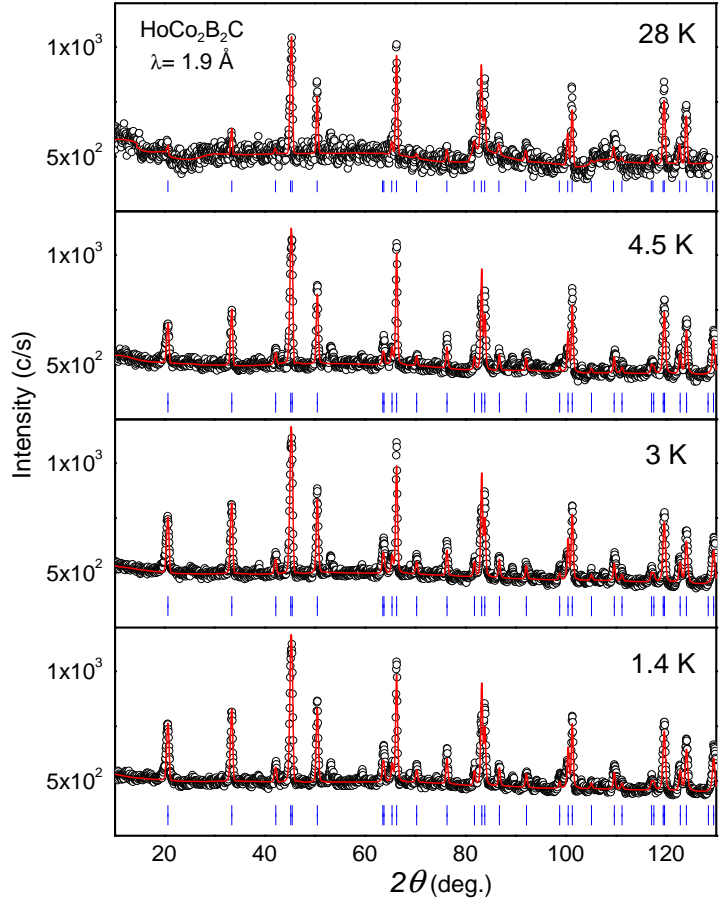}%
\caption{(Color online) Rietveld refinement of representative powder
diffractograms of HoCo$_{2}$B$_{2}$C. Symbols, solid line, and short vertical
bars represent, respectively, measured intensities, calculated intensities
based on Rietveld-refinement, and positions of the Bragg reflections. The
total contribution is composed of nuclear peaks (\textit{I4/mmm}) and the FM
order of the Ho subsystem (see text).}%
\label{hoco2b2c-nd-riet}%
\end{center}
\end{figure}

\begin{figure}
[th]
\begin{center}
\includegraphics[
height=4.8136in,
width=3.7265in
]%
{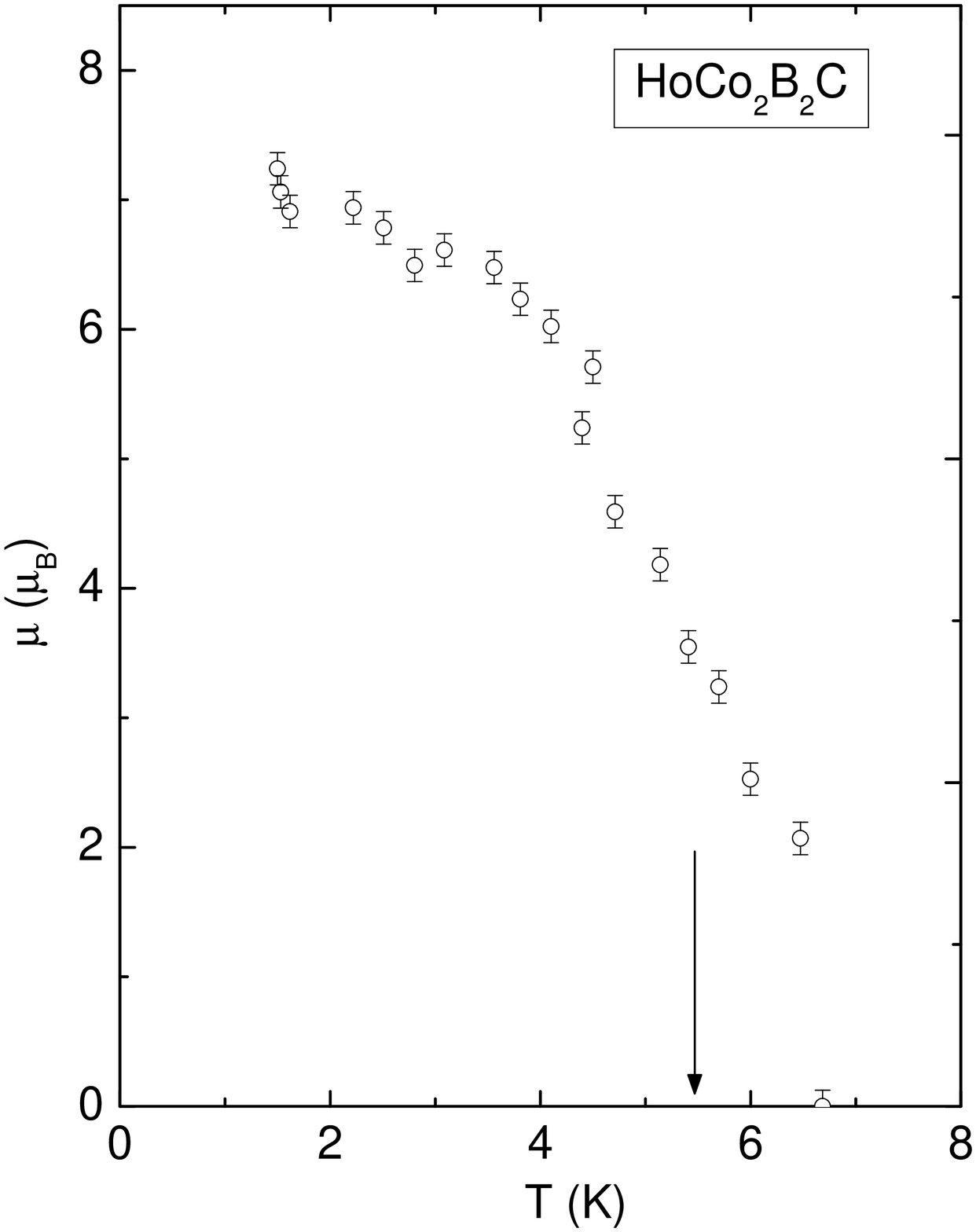}%
\caption{Thermal evolution of the ordered magnetic moment for HoCo$_{2}$%
B$_{2}$C as obtained from the neutron diffraction refinements of Fig.
\ref{hoco2b2c-nd-riet}.}%
\label{hoco2b2c-moment}%
\end{center}
\end{figure}

The diffractograms of \textrm{HoCo}$_{2}$\textrm{B}$_{2}$\textrm{C} (Fig.
\ref{hoco2b2c-raw}) show that there are no additional satellite peaks when the
sample is cooled to below $T_{C}$ = 5.4(2) K; instead, there is a considerable
enhancement in the intensities of the fundamental Bragg peaks. The resulting
magnetic pattern is easily indexed on the basis of a FM unit cell which is of
the same dimension as the crystalline one. The Rietveld analysis (see Fig.
\ref{hoco2b2c-nd-riet}) confirms the conclusions drawn from the magnetic
indexing: a FM structure with the moments lying within the basal plane. The
thermal evolution of the Ho$^{3+}$ moments (Fig. \ref{hoco2b2c-moment}) shows
a tendency towards saturation at an ordered value of the moment of 7.2(2)
$\mu_{B}$. This is in excellent agreement with the saturation moment of 7.3(1)
$\mu_{B}$ obtained from the isothermal magnetization measurement (see
Fig.\ref{Fig-rco2b2c-magnetization}). Furthermore, Fig \ref{hoco2b2c-moment}
indicates that the intensity monotonically decreases as $T_{C}$ is being
approached from below, and at higher T decreases almost linearly and becomes
zero at 7.6 K. The presence of magnetic intensity well above $T_{C}$
(attributed to short range order) has already been observed in the specific
heat and susceptibility of \textrm{HoCo}$_{2}$\textrm{B}$_{2}$\textrm{C.}%
\cite{00-RCo2B2C} Similar short-range features were observed in the neutron
diffraction studies on the \textrm{HoNi}$_{2}$\textrm{B}$_{2}$\textrm{C}
isomorph.\cite{Lynn97-RNi2B2C-ND-mag-crys-structure}%

%

%

In stark contrast with the exotic features of the low-temperature magnetic
phase diagram of \textrm{HoNi}$_{2}$\textrm{B}$_{2}$\textrm{C}%
,\cite{Muller01-interplay-review,Canfield-RNi2B2C-Hc2-review} the above
results demonstrate that \textrm{HoCo}$_{2}$\textrm{B}$_{2}$\textrm{C} orders
into a simple FM\ state and that within the studied temperature range [1.4
K$<T<T_{C}$] there are no manifestations of additional zero-field
order-to-order magnetic transformations. This corrects our earlier report of a
magnetic transition at $T_{m}$=1.5 K.\cite{00-RCo2B2C} Even though in this
work we were not able to collect diffractograms below 1.4 K, considering the
similarity with \textrm{ErCo}$_{2}$\textrm{B}$_{2}$\textrm{C} (see \S \ IV.C)
and \textrm{TbCo}$_{2}$\textrm{B}$_{2}$\textrm{C}
(Ref.\cite{08-Mag-Structure-TbCo2B2C}), no order-to-order transitions or
modifications in the magnetic structure of Ho sublattice are expected to occur.

\subsection{DyCo$_{2}$B$_{2}$C}

\begin{figure}
[th]
\begin{center}
\includegraphics[
height=4.6855in,
width=3.474in
]%
{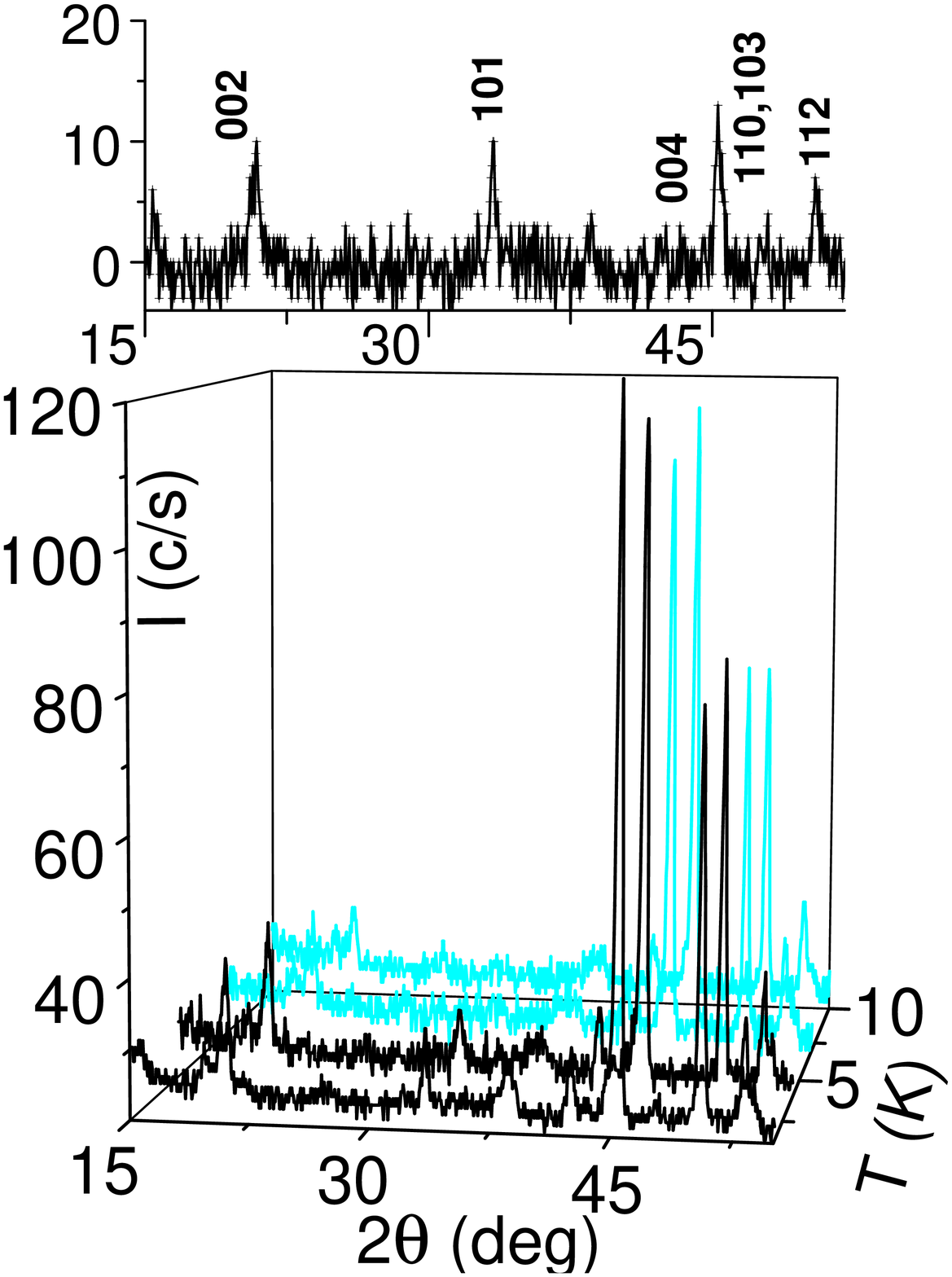}%
\caption{(Color online) Lower panel: representative diffractograms of
\ \textrm{DyCo}$_{2}$\textrm{B}$_{2}$\textrm{C } measured at 1.4, 4, 7, and 10
K. The upper panel shows the difference plot (after subtracting of the pattern
at 10 K from the one at 1.4 K): the pattern is indexed based on a
ferromagnetic structure.}%
\label{dyco2b2c-raw}%
\end{center}
\end{figure}

Figure \ref{dyco2b2c-raw} shows the thermal evolution of the diffraction
pattern of \textrm{DyCo}$_{2}$\textrm{B}$_{2}$\textrm{C} when cooled through
the magnetic transition temperature [$T_{C}=$ 8.0(2) K].\cite{00-RCo2B2C} On
subtracting the paramagnetic diffractogram at 10 K from the one at 1.4 K, we
obtained the magnetic diffraction pattern shown in the inset of Fig.
\ref{dyco2b2c-raw}. This pattern is indexed on the basis of a FM cell having
the same cell dimensions as those of the crystalline one. We attempted to
refine the neutron-diffractogram (Fig. \ref{dyco2b2c-raw}), but due to the
relatively high neutron absorption of Dy, the statistics are not adequate to
allow a reliable determination of the ordered moment. Nevertheless, it is
inferred that the moments lie within the $ab$-plane and its value at least
exceeds 4.6 $\mu_{B}$. To get a better evaluation of the saturated moment, we
resorted to the magnetization isotherm. Fig. \ref{Fig-rco2b2c-magnetization}
indicates that the saturated moment at 1.35 K is 6.2(1) $\mu_{B}$; this value
is 29\% lower than the value reported for the \textrm{DyNi}$_{2}$%
\textrm{B}$_{2}$\textrm{C} isomorph.
%

\begin{figure}
[th]
\begin{center}
\includegraphics[
height=4.811in,
width=3.7031in
]%
{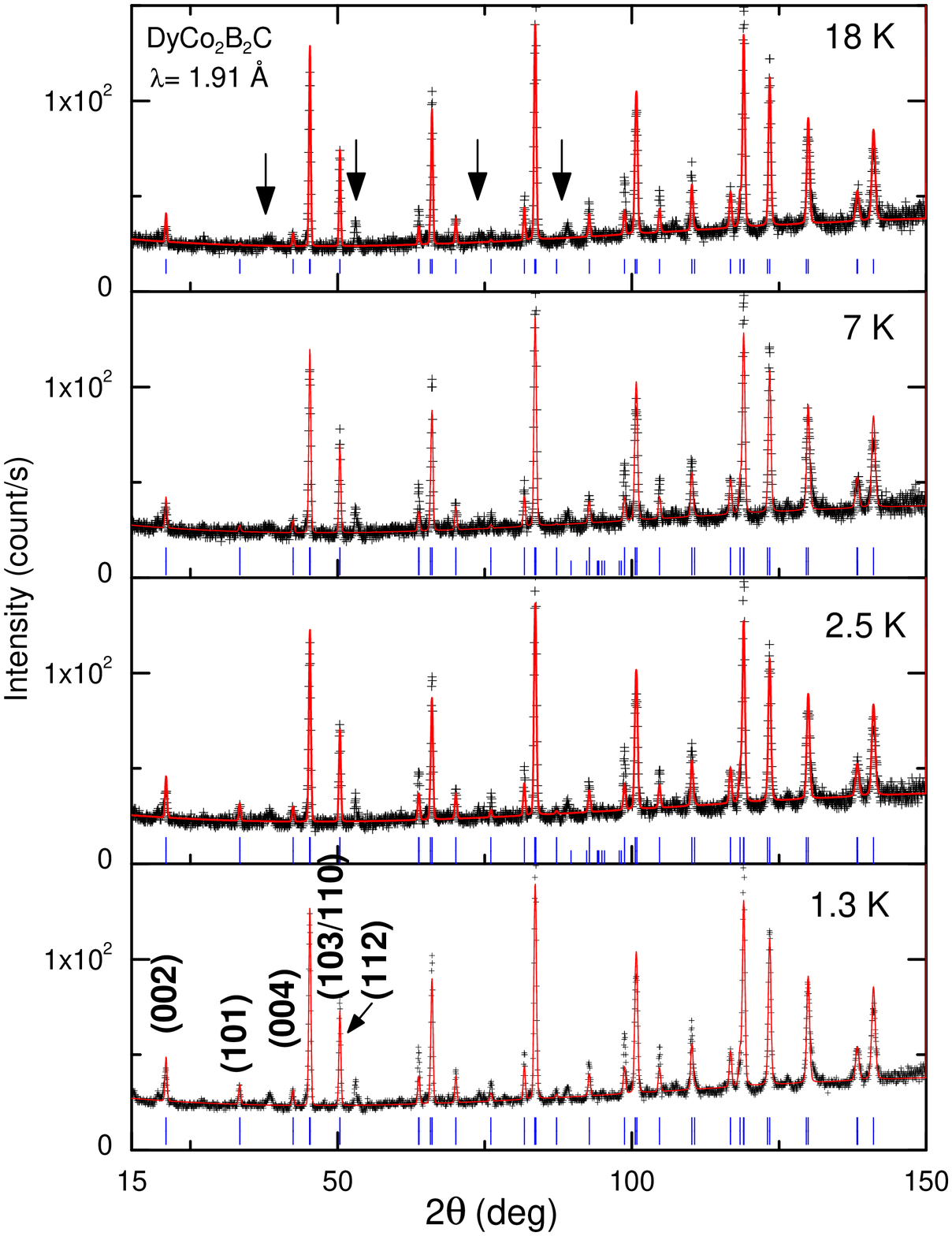}%
\caption{Refinements of representative diffractograms of \textrm{DyCo}$_{2}%
$\textrm{B}$_{2}$\textrm{C}. Except for the weak impurity peaks (denoted by
short vertical arrows), all peaks (symbol) are fit (solid lines) based on the
following crystalline and magnetic structures: a tetragonal unit cell
($I$4$/mmm$ with lattice parameters as given in Table I) and a FM unit cell
(Bragg peaks shown as short vertical bars) having the same dimensions.}%
\label{Dyco2b2c-nd-riet}%
\end{center}
\end{figure}

\section{Discussion and Conclusion}

As mentioned above, the crystalline electric field environment around the
$R^{3+}$ site in the $RT_{2}$\textrm{B}$_{2}$\textrm{C}\ family is expected to
be the same as the one observed in the isomorphous $R$\textrm{Ni}$_{2}%
$\textrm{B}$_{2}$\textrm{C }systems. Consequently the single-ion properties of
$R^{3+}$ (such as the ordered moment, direction, and magnetoelastic
properties) should be similar. In particular, the magnetic easy axes in the
sequence of compounds \textrm{Er}$T_{2}$\textrm{B}$_{2}$\textrm{C},
\textrm{Ho}$T_{2}$\textrm{B}$_{2}$\textrm{C}, and $\mathrm{Dy}T_{2}$%
\textrm{B}$_{2}$\textrm{C} ($T$\ =Co, Ni) should be similar (see Table
\ref{Tab.I}). This working assumption is\ particularly helpful since the
zero-field powder neutron diffraction technique is not able to identify the
easy axis, even though it successfully identified the easy character of the
$ab$-plane. Then based on this similarity of the CEF properties, the easy axis
of both $R$= Ho, \textrm{Dy} is taken to be ($1,1,0$) while that of $R$= Er is
($1,0,0$) axis. It is reassuring that these arguments proved to be valid for
the case of \textrm{TbCo}$_{2}$\textrm{B}$_{2}$\textrm{C }(see Table
\ref{Tab.I} and Ref. \cite{08-Mag-Structure-TbCo2B2C}).\begin{table}[th]
\caption{Some structural and magnetic parameters of the isomorphous
$R$\textrm{Ni}$_{2}$\textrm{B}$_{2}$\textrm{C} and $R$\textrm{Co}$_{2}%
$\textrm{B}$_{2}$\textrm{C }series\textrm{ }( \textit{R }= Tm, Er, Ho, Dy,
Tb). The room-temperature cell dimensions of $R$\textrm{Ni}$_{2}$%
\textrm{B}$_{2}$\textrm{C} ($R$\textrm{Co}$_{2}$\textrm{B}$_{2}$\textrm{C})
are determined from neutron\cite{Lynn96-RNi2B2C-mag-structure}
(X-ray\cite{00-RCo2B2C,08-Mag-Structure-TmCo2B2C}) diffraction. The magnetic
properties of $R$\textrm{Ni}$_{2}$\textrm{B}$_{2}$\textrm{C} are taken from
Ref.\cite{Lynn96-RNi2B2C-mag-structure,Chang96-TmNi2B2C-mag-struct} while
those of \textrm{TbCo}$_{2}$\textrm{B}$_{2}$\textrm{C} and \textrm{TmCo}$_{2}%
$\textrm{B}$_{2}$\textrm{C} are taken, respectively, from Ref.
\cite{08-Mag-Structure-TbCo2B2C} and Ref. \cite{08-Mag-Structure-TmCo2B2C}.
The symbols have their usual meanings. $\left\vert \vec{\mu}\right\vert $
refers to the value obtained from the neutron diffraction analysis.}%
\begin{tabular}
[c]{ccccccccccc}\hline\hline
$R$ & \multicolumn{2}{c}{Tm} & \multicolumn{2}{c}{Er} & \multicolumn{2}{c}{Ho}
& \multicolumn{2}{c}{Dy} & \multicolumn{2}{c}{Tb}\\\hline
$M$ & Co & Ni & Co & Ni & Co & Ni & Co & Ni & Co & Ni\\\hline
a ($\operatorname{\mathring{A}}$) & 3.473 & 3.4866 & 3.4844\qquad & 3.5019 &
3.4997\qquad & 3.5177 & 3.5099\qquad & 3.53420 & 3.5246\qquad & 3.5536\\
c ($\operatorname{\mathring{A}}$) & 10.647 & 10.607 & 10.5902 & 10.5580 &
10.54465 & 10.5278 & 10.5268 & 10.4878 & 10.5176 & 10.4352\\
$T_{crit}$ (K) & 0.8 & 1.53 & 4.0 & 6.8 & 5.4 & 5.0 & 8.0 & 10.6 & 6.3 &
15.0\\
Mode & FM & TSW & AFM & TSW & FM & AF & FM & AF & FM & LSW\\
$k$ & (0,0,0) & (0.093,0.093,0) & ($\frac{1}{2},0,\frac{1}{2}$) &
(0.553,0,0) & (0,0,0) & (0,0,1) & (0,0,0) & (0,0,1) & (0,0,0) & (0.555,0,0)\\
$\left\vert \vec{\mu}\right\vert $ & \symbol{126}1 & 3.8 & 6.8(2) & 7.2 &
7.2(2) & 8.6 & $>$4.6 & 8.5 & 7.6 & 7.8\\
easy axis & (0,0,1) & (0,0,1) & (0,1,0) & (0,1,0) & (1,1,0) & (1,1,0) &
(1,1,0) & (1,1,0) & (1,0,0) & (1,0,0)\\\hline\hline
\end{tabular}
\label{Tab.I}%
\end{table}

The observation that none of the studied $R$\textrm{Co}$_{2}$\textrm{B}$_{2}%
$\textrm{C} compounds exhibit an incommensurate modulated state is argued to
be an indication that the Fermi surface nesting features are
quenched.\cite{08-Mag-Structure-TbCo2B2C} Another striking difference between
the isomorphous $RT_{2}$\textrm{B}$_{2}$\textrm{C} ($R$=Dy, Ho) compounds is
shown in Fig. \ref{Fig-rco2b2c-magnetization}: the absence of metamagnetic
field-induced transitions in the magnetization isotherms. This feature is most
evident in the\ magnetization isotherms of single-crystal \textrm{TbCo}$_{2}%
$\textrm{B}$_{2}$C\textrm{ }(see Fig. 2 of Ref.
\cite{08-Mag-Structure-TbCo2B2C}). These two observations highlight the
delicate balance of the exchange, crystalline, and dipolar interactions in
$R$\textrm{Ni}$_{2}$\textrm{B}$_{2}$\textrm{C} for establishing their exotic
magnetic phase diagrams: it is recalled that \textrm{GdNi}$_{2}$%
\textrm{B}$_{2}$\textrm{C} (wherein there are exchange and dipolar forces but
no CEF interaction) does manifest a modulated state but no cascade of
field-induced metamagnetic transformations. On the other hand, none of the
studied $R$\textrm{Co}$_{2}$\textrm{B}$_{2}$\textrm{C} ($R$=Dy, Ho) compounds
exhibit a modulated state or a cascade of metamagnetic transformations even
though these compounds do possess exchange, CEF and dipolar forces that are
similar to those of the Ni-based isomorphs.

The FM structure of $R$\textrm{Co}$_{2}$\textrm{B}$_{2}$\textrm{C} ($R$= Tm
[Ref. \cite{08-Mag-Structure-TmCo2B2C}], Ho, Dy, Tb [Ref.
\cite{08-Mag-Structure-TbCo2B2C}]) is expected to generate a large internal
molecular field at the Co-sublattice. Then if this internal field exceeds a
certain critical value, the Co-sublattice could be spontaneously
polarized.\cite{Bloch-Lemaire70-RCo2,Bloch75-RCo2,Cyrot79-electr-cal,Cyrot79-b-electr-cal}
We were particularly interested in this prospect of Co-sublattice polarization
since this scenario was thought to support our earlier report of a second
metamagnetic transformation ($T_{M}$ boundary) in the magnetic phase diagram
of $R$\textrm{Co}$_{2}$\textrm{B}$_{2}$\textrm{C}.\cite{00-RCo2B2C} It
happened that the neutron diffractograms of $R$=Dy, Ho, Er compounds as well
those of \textrm{TbCo}$_{2}$\textrm{B}$_{2}$\textrm{C} [Ref.
\cite{08-Mag-Structure-TbCo2B2C}] and \textrm{TmCo}$_{2}$\textrm{B}$_{2}%
$\textrm{C} [Ref. \cite{08-Mag-Structure-TmCo2B2C}] do not show any intrinsic
transformation. Furthermore, this $T_{M}$ transition is not always manifested
in the recent thermodynamical measurements on \textrm{TbCo}$_{2}$%
\textrm{B}$_{2}$\textrm{C}.\cite{08-Mag-Structure-TbCo2B2C} Consequently, it
is concluded that there is no\ spontaneous polarization of the Co-sublattice:
if there is any it must be sample dependent.

In conclusion, this work showed that the ground states as well as the magnetic
phase diagrams of the studied $R$Co$_{2}$B$_{2}$C compounds are distinctly
different from those of their Ni-based isomorphs. These findings should
certainly contribute to our understanding of the magnetism, superconductivity
and their interplay in the borocarbides series.

\begin{acknowledgments}
We acknowledge the partial financial support from the Brazilian agencies CNPq
(485058/2006-5), Faperj (E-26/171.343/2005), and FAPESP (2008/00457-2).
\end{acknowledgments}

\bibliographystyle{apsrev}
\bibliography{Borocarbides,Crystalograph,Intermetallic,Mag-classic,Massalami,ND-RepAnalysis,To-Be-Published}

\end{document}